\documentclass[preprint]{aastex}
\usepackage{color}
\usepackage{subfigure}
\usepackage{epsfig}
\usepackage{lscape}
\usepackage{soul}
\usepackage[normalem]{ulem}

\shorttitle{C/O of 55 Cnc}
\shortauthors{Teske et al.}

\begin{document}

\newcommand{\txw}{\textwidth}
\title{Carbon and Oxygen Abundances in Cool Metal-rich Exoplanet
  Hosts: A Case Study of the C/O Ratio of 55 Cancri{*}}

\altaffiltext{*}{The data presented herein were obtained at the W.M. Keck Observatory, which is operated as a scientific partnership among the California Institute of Technology, the University of California and the National Aeronautics and Space Administration. The Observatory was made possible by the generous financial support of the W.M. Keck Foundation.}

\author{Johanna K. Teske\altaffilmark{1}, Katia Cunha\altaffilmark{1,
    2}, , Simon C. Schuler\altaffilmark{3}, Caitlin
  A. Griffith\altaffilmark{4}, Verne V. Smith\altaffilmark{5}}

\altaffiltext{1}{Steward Observatory, University of Arizona, Tucson, AZ, 85721, USA; email: jteske@as.arizona.edu}
\altaffiltext{2}{Observat\'orio Nacional, Rua General Jos\'e Cristino,
  77, 20921-400, S\~ao Crist\'ov\~ao, Rio de Janeiro, RJ, Brazil}
\altaffiltext{3}{University of Tampa, 401 W. Kennedy Blvd., Tampa, FL 33606, USA}
\altaffiltext{4}{Lunar and Planetary Laboratory, University of Arizona, Tucson, AZ, 85721, USA}
\altaffiltext{5}{National Optical Astronomy Observatory, 950 North
  Cherry Avenue, Tucson, AZ 85719, USA}


\begin{abstract}
The super-Earth exoplanet 55 Cnc e, the smallest member of a
five-planet system, has recently been observed to transit its host
star. The radius estimates from transit observations, coupled with
spectroscopic determinations of mass, provide constraints on its
interior composition.
The composition of exoplanetary interiors and
atmospheres are particularly sensitive to elemental C/O ratio, which to
first order can be estimated from the host stars. 
Results from a recent spectroscopic study analyzing the
6300\,{\AA} [O I] line and two C I lines suggest that 55 Cnc has a
carbon-rich composition (C/O=1.12$\pm$0.09). 
However oxygen abundances derived using the 6300\,{\AA} [O I] line are highly
sensitive to a Ni I blend, particularly in metal-rich stars such as 55
Cnc ([Fe/H]$=$0.34$\pm$0.18). 
Here, we further investigate 55 Cnc's
composition by deriving the carbon and oxygen abundances from these
and additional C and O absorption features. We find that the measured C/O
ratio depends on the oxygen lines used. The C/O ratio that we derive
based on the 6300\,{\AA} [O I] line alone is consistent with the
previous value. Yet, our investigation of additional abundance
indicators results in a mean C/O ratio of 0.78$\pm$0.08. The lower
C/O ratio of 55 Cnc determined here may place this system at the
sensitive boundary between protoplanetary disk compositions giving
rise to planets with high ($>$0.8) versus low ($<$0.8) C/O ratios.
This study illustrates the caution that must applied when determining
planet host star C/O ratios, particularly in cool, metal-rich stars.

\end{abstract}


\section{Introduction}
Exoplanet observational surveys reveal a large and diverse population of
planets with masses between a few and $\sim$20 Earth masses,
approaching the size of Solar System terrestrial planets (Lovis et
al.\,2009; Sumi et al.\,2010; Borucki et al.\,2011). 
A member of the five-planet system orbiting 
a nearby ($\sim$12.3 pc) G8V star every 18 hours, 55
Cnc e (e.g., McArthur et al.\,2004; Winn et al.\,2011)
belongs to the small sample of confirmed 
terrestrial-sized planets that transit their host stars. 
Observations of 55 Cnc e have provided a well-constrained mass
(8.37$\pm$0.38 M$_{\oplus}$; Endl et al.\,2012) 
and radius (e.g.,
1.990$^{+0.084}_{-0.080}$ R$_{\oplus}$ in the visible; Dragomir et
al.\,2013), yielding the density of the super-Earth exoplanet (5.86$^{+0.79}_{-0.76}$ g cm$^{-3}$)
, which can then be used to constrain its interior composition. 

The observed mass and radius of 55 Cnc e place it between the
high-density ``super-Mercuries'', like CoRoT-7b and Kepler-10b, and the
volatile-rich small planets, like Kepler-11b and GJ 1214b. 
It intersects the threshold mass and radius between interior compositions
that necessarily require volatiles and ones that may be rocky (see,
for example, Gillon et al.\,2012, Figure 5). 
Hence a massive water envelope
($\simeq10$\%), which would be super-critical given 55 Cnc e's
irradiation, 
over an Earth-like interior (33\% iron core
above 67\% silicate mantle with 10\% iron by mol), has been suggested
to explain the observed mass and radius 
(Winn et al.\,2011; Demory et al.\,2011; Gillon
et al.\,2012). 

Recently Madhusudhan et al.\,(2012) suggest an alternative and
carbon-rich composition of 55 Cnc e, garnering the super-Earth popular
attention as ``the diamond planet.''
Measurements of the carbon and oxygen abundances from two C I lines
(5052\,{\AA}, 5135\,{\AA}) and one forbidden [O I] line (6300\,{\AA})
indicate a C/O\footnote{The C/O ratio -- the ratio of the number of carbon atoms
   to oxygen atoms -- is calculated in stellar abundance analysis as C/O=
  N$_{\rm{C}}$/N$_{\rm{O}}$=10$^{\rm{logN(C)}}/10^{\rm{logN(O)}}$ where
 log(N$_{\rm{X}}$)=log$_{10}$(N$_{\rm{X}}$/N$_{\rm{H}}$)+12.} ratio of
1.12$\pm$0.19 (Delgado Mena et al.\,2010), i.e., a highly carbon-rich
star compared to the solar C/O$\sim$0.50 (Asplund et al.\,2005). If the disk shared the host star's composition, and the host star is
carbon-rich, then the planetesimals accreted during the formation of
55 Cnc e were likely Fe- and C-rich (Bond et al.\,2010; Madhusudhan et
al.\,2012).
To investigate the composition of the possibly carbon-rich exoplanet,
Madhusudhan et al.\,(2012) consider two families of carbon-rich
interior models of 55 Cnc e, consisting of layers, from inner to
outer, of Fe-SiC-C and Fe-MgSiO$_{3}$-C. 
Included in their carbon equation of state (EOS) are the graphite EOS at low pressures, the phase transition to
diamond between 10 GPa$<$P$<$1000 GPa, and the Thomas-Fermi-Dirac EOS
at high pressures. Madhusudhan et al.\,(2012) find a wide range
of compositions are possible, including extreme combinations like (Fe,
SiC, C) = (33\%, 0\%, 67\%), and the best match to 55 Cnc e's
observations depends on the adopted radius measurement, and the conditions
in the protoplanetary disk, e.g. temperature, at which the building
blocks of the planet condense. 


The exact composition of 55 Cnc e
depends on the primary source of accreted planetesimals, 
the ratio of gas to solid
material accreted, and how isolated the atmosphere was from the
interior (e.g., {\"O}berg et al.\,2011; Bond et al.\,2010). While the
C/O ratios of protoplanetary disks likely change with time and
distance ({\"O}berg et al.\,2011), the assumption that the disk bears
roughly the same composition of the host star is a reasonable first-order one for
estimating refractory condensates forming rocky planets (Bond et
al.\,2010; Carter-Bond et al.\,2012; Johnson et al.\,2012). Thus
constraining the elemental abundances of the host star is a crucial step in determining the composition of 55 Cnc e. 

Yet determinations of the stellar C/O ratios can be challenging. The
6300\,{\AA} forbidden oxygen line is chosen in many studies, including
the previous study of 55 Cnc, because of it has been shown to give
reliable abundances in LTE analyses (e.g., Schuler et al.\,2011; Cuhna
et al.\,1998). However,
this line is weak and blended with a Ni I line, the treatment of which
significantly affects the derived oxygen abundance, particularly at
high metallicities. Here we further investigate the C/O ratio of 55
Cnc by determining the nickel abundance from the data and reanalyzing
the original line used to study its oxygen content, as well as the
same two C I lines. We also
determine the oxygen abundance from an additional forbidden [O I] line
at 6363\,{\AA}
and the O I triplet at 7774\,{\AA}, and the carbon abundance from two molecular
C$_{2}$ features. This work aims to determine whether the stellar
abundance indicates a diamond-rich composition of 55 Cnc, and to
explore the difficulties in deriving the C/O ratios in
cool high-metallicity stars.

\section{Observations and Abundance Analysis} 
\subsection{Data}
We analyze Keck/HIRES (Vogt et al.\,1994) archive spectra of 55 Cnc (PID H32bH; PI
Shkolnik) taken across four nights in January 2006, covering the
wavelength range 3360-8100\,{\AA} with the kv370 filter. Individual frame exposure
times range from 20 to 120 sec and S/N ratios range from $\sim$170 to 350 around the
6300\,{\AA} [O I] line; the 35 spectra combined yield a S/N of
$\sim$1270 around the 6300\,{\AA} [O I] line. To enable differential abundance
determinations relative to the Sun, we also analyze three
solar spectra of
reflected light from Vesta (PID N014Hr; PI Marcy). These data were taken in
April 2006 with the same filter, and with individual frame exposure
times $\sim$230 sec; combined the spectra yield a S/N ratio $\sim$315 around the
6300\,{\AA} [O I] line. 
All archive HIRES data were reduced with the MAKEE
pipeline\footnote{www.astro.caltech.edu/~tb/makee/} using
corresponding bias ($\sim$3), flat ($\sim$30), ThAr (arc), and trace star frames for each target frame
separately. The frames were then combined in IRAF\footnote{IRAF
  is distributed by the National Optical Astronomy Observatory, which
  is operated by the Association of Universities for Research in
  Astronomy (AURA) under cooperative agreement with the National
  Science Foundation.}.  

\subsection{Stellar Parameters}
The stellar parameters ($T_{\rm{eff}}$, log $g$,
microturbulence [$\xi$]) and metallicity
([Fe/H]{\footnote{[X/H]=log(N$_{X}$) - log (N$_X$)$_{solar}$}}) for 55 Cnc were derived
following the procedures in Schuler et al.\,(2011) and Teske et
al.\,(2013). We measured equivalent widths (EWs) of 55 Fe I lines and 9 Fe II lines in 55
Cnc and the Sun (with the one-dimensional spectrum analysis
package SPECTRE; Fitzpatrick \& Sneden 1987). We fit Gaussian profiles to each
absorption line (some weaker lines were fit with a
Simpson's Rule integration). 

The abundances were determined using
an updated version of the 
LTE spectral analysis code MOOG (Sneden 1973), with model atmospheres
interpolated from the Kurucz ATLAS9 grids\footnote{See http://kurucz.harvard.edu/grids.html}. 
To fulfill the requirement of excitation equilibrium, 
the [Fe/H] values
derived from the Fe I lines must not show any correlation with the
lower level excitation potential ($\chi$); this was used to determine $T_{\rm{eff}}$.  
In addition, the $\xi$ was determined by requiring [Fe/H] values derived from the Fe I lines
to show no correlation with 
the measured EW values (specifically the
reduced equivalent width, log(EW/$\lambda$)). 
Also, 
the averaged [Fe/H]
values derived from the Fe I and Fe II lines must be equal -- the requirement
of ionization equilibrium; this sets the surface gravity (log $g$). 

Initial values of $T_{\rm{eff}}$, log $g$, microturbulence ($\xi$),
and [Fe/H] of 55 Cnc from the literature were taken as starting values in
the iterative process of determining 55 Cnc's stellar parameters. 
Prior to this
iterative scheme, we 
ensured that there was no correlation between $\chi$ and the reduced EWs of the Fe I
lines analyzed; unique solutions for $T_{\rm{eff}}$ and $\xi$ are only
possible if there is no such correlation. The measured reduced EWs
were used to determine 
abundances (using the``abfind'' task in MOOG), and the
stellar parameters were altered and new [Fe/H] abundances determined
until the criteria 
above were met.  
The log$N$(Fe) values from each line were normalized to solar values on
a line-by-line basis. 
The log$N$(Fe) value for
the Sun was determined with our solar spectrum and a solar Kurucz model with $T_{\rm{eff}}$=5777, log $g$=4.44, [Fe/H]=0.00, and $\xi$=1.38.

\subsubsection{Uncertainties in $T_{\rm{eff}}$, log $g$, and
$\xi$}
The errors in $T_{\rm{eff}}$ and
$\xi$ were calculated by forcing 1$\sigma$ correlations in the relations
between [Fe I/H] and $\chi$, and between [Fe I/H] and reduced
EW, respectively. The change in $T_{\rm{eff}}$ or
$\xi$ required to cause a correlation coefficient $r$ significant
at the 1$\sigma$ level was adopted as the uncertainty in these
parameters. The uncertainty in log $g$ was calculated differently,
through an iterative process described in detail in Baubar \&
King\,(2010). The difference between the Fe I and Fe II abundances is
dependent on the log $g$ value, so the uncertainty in log $g$ is tied
to the uncertainties in both [Fe I/H] and [Fe II/H]. To calculate the
uncertainty in log $g$, its value is perturbed until the difference
between [Fe I/H] and [Fe II/H] is equal to the combined uncertainty in
[Fe I/H] and [Fe II/H]. Uncertainties in [Fe I/H] and [Fe II/H] are calculated from the quadratic sum of the
individual uncertainties in these abundances due to the derived
uncertainties in $T_{\rm{eff}}$ and $\xi$ as well as the uncertainty
in the mean ($\sigma_{\mu}$\footnote{$\sigma_{\mu}=\sigma/\sqrt{N-1}$,
where $\sigma$ is the standard deviation of the derived abundances and
$N$ is the number of lines used to derive the abundance.}) of each
abundance (see \S2.3.3). The same procedure is then repeated,
including the first iteration's log $g$ uncertainty (the $\delta$log
$g$) in the calculation of the Fe abundance uncertainties. The final
log $g$ uncertainty is then the difference between the log $g$ value
originally derived 
and that obtained from this second iteration of the error calculation. 

Table \ref{tab:stellar_params} lists the final derived stellar
parameters and 1$\sigma$ uncertainties, as well as several
literature values for comparison. 
The errors derived here are larger than those from
previous studies, in which 55 Cnc was part of a large ensemble of
stars analyzed. 
We note, however, that using the same stellar
parameter analysis and error
calculation method on stars with temperatures closer to the
  Sun than 55 Cnc,
we obtain
errors on $T_{\rm{eff}}$, log $g$, and $\xi$ that are more similar to
typical error values quoted in the literature. 
Larger errors for cooler stars are also found by Teske et
al.\,(2013), Ammler-von Eiff et al.\,(2009), and Torres et
al.\,(2012), studies that determine stellar parameters and errors with methods
similar to those used in this work.
Our conservative stellar
parameter errors for 55 Cnc also propagate
through the abundance errors, as discussed in \S2.3.3. 

\subsection{Stellar Abundances}

Abundances of iron and nickel ([Fe/H], [Ni/H]) were
normalized to solar values on a line-by-line basis, 
derived directly from EW measurements of
spectral lines in 55 Cnc and the Sun 
with the ``abfind'' driver in MOOG. 
Lines lists for Fe and Ni are from Schuler et
al.\,(2011), and lower level excitation potentials ($\chi$)
and transition probabilities (log $gf$) are taken from the Vienna
Atomic Line Database (VALD; Kupka et al. 1999), although we note that the log $gf$ values do
not have a impact on the final abundances due to our
strictly differential analysis. The EW measurements (and results of
our synthesis analysis, described below) are shown in
Table \ref{tab:lines}, along with the wavelength, $\chi$, log $gf$,
EWs, and line-by-line abundances for each element for the Sun and 55
Cnc. 
 

\subsubsection{Carbon Abundance}
The carbon abundance for 55 Cnc was derived from two C I lines at
5052\,{\AA} and 5380\,{\AA} and two C$_2$ molecular features at
5086.3\,{\AA} and 5135.6\,{\AA}. The two
C I lines have been shown to provide reliable abundances in
solar-type stars, with negligible NLTE corrections ($\leq$0.05 dex;
Asplund et al.\,2005; Takeda \& Honda 2005; Caffau et al.\,2010). We derived [C/H] from these
lines with our EW measurements, 
with atomic parameters from Hibbert et
al.\,(1993) (see Table \ref{tab:lines}). 
The log$N$(C)$_{\odot}$ values we derive with our EW measurements are
a good match, with $\leq$0.02 dex difference, to the log$N$(C)$_{\odot}$ values derived by Caffau et al.\,(2010) from
these lines using 3D hydrodynamical simulations of the Sun. 

The C$_2$ lines are blends of
multiple components of the Swan system, requiring spectral synthesis (matching a set of trial synthetic spectra to
the observed spectrum)
for abundance derivation. 
We used
the line lists of Schuler et al.\,(2011) 
and C$_2$ molecular data from Lambert \& Ries (1981),
modified in that paper from theoretical values to fit the Kurucz solar flux atlas assuming a solar abundance
of log$N$(C)$_{\odot}$=8.39 (Asplund et al.\,2005). A dissociation energy
of 6.297 eV was assumed for C$_2$ (Urdahl et al.\,1991). The
synthesized spectra were convolved with a Gaussian profile, based on
near-by unblended lines, to represent the instrument
PSF, stellar macroturbulence, and rotational broadening; the remaining free parameters were
continuum normalization, line broadening, wavelength shift, and carbon abundance. The best fits to the synthesized
spectra for the C$_2$ lines were determined by minimizing the
deviations between the observed and synthetic spectra. 

As evidenced in
Table \ref{tab:choh}, the [C/H] abundance derived from the C I is
slightly lower than that derived from the
C$_2$ lines; this as also observed by Asplund et
al.\,(2005) in both 3D hydrodynamical and 1D models of the solar
atmosphere. However, our [C/H]$_{\rm{C\,I}}$ value overlaps with the [C/H]$_{\rm{C_{2}}}$
value within errors. 

\subsubsection{Oxygen Abundance}
Oxygen abundances were derived from three separate indicators and are
listed in Table \ref{tab:choh}. The forbidden [O I] line at 6300.3\,{\AA} is well-described by LTE (e.g. Takeda
2003). This line is blended with a Ni I line (2 isotopic
components) with a strength $\sim$55\% of the [O I] line in the Sun (Caffau
et al. 2008), requiring spectral synthesis similar to the C$_2$
lines. Due to [Ni/Fe] increasing with [Fe/H] (Bensby et al.\,2003),
the Ni I blend becomes more important at higher metallicities, the
regime in which most high-C/O values for exoplanet host stars have been found (Nissen 2013). When determining the oxygen abundance, we used the
nickel abundance measured directly from our 55 Cnc spectrum, log$N$(Ni) =
6.68 derived from 14 lines, with log $gf$($^{60}$Ni)=-2.965 and log
$gf$($^{58}$Ni)=-2.275 (Bensby et al.\,2004). For the 6300.3\,{\AA}
line we adopted the Storey \&
Zeippen\,(2000) log $gf=$-9.717 value, based
on their forbidden transition probability calculations including both
relativistically-corrected magnetic dipole and electric quadruopole
contributions. 

The [O I] 6363.79\,{\AA} forbidden line (log $gf=-$10.185,
Storey \& Zeippen\,2000) 
is weaker than the 6300\,{\AA} line, and is also
blended with CN lines (6363.776\,{\AA} and
6363.846\,{\AA}; Asplund et al. 2004). 
We again
determined [O/H] from this line using spectral synthesis, with a line list
compiled mostly from Kurucz\footnote{http://kurucz.harvard.edu} and
supplemented with lines from Asplund et al.\,(2004). For this
analysis, we used the carbon abundance derived here, and a
solar-scaled nitrogen abundance. 

The O I triplet lines at 7771-7775\,{\AA} are unblended and
prominent, hence we analyzed them with direct EW
measurements (see Table \ref{tab:lines}).  
Here we
did not include the 7774\,{\AA} component because 
the line appears slightly asymmetric, and gives 
an anomalously high ($+$0.10 dex) abundance compared to the other two
components, 7771.94\,{\AA} ($\chi$=9.15 eV, log $gf$=0.369;
Hibbert et al.\,1991) and 7775.4\,{\AA} ($\chi$=9.15 eV,
log $gf$= 0.001; Hibbert et al.\,1991). This 
effect is also seen in the coolest stars in Schuler et al.\,(2006) and Bubar \& King
(2010); these authors suggest it may be due to a Fe I blend at
7774.00\,{\AA} in cool metal-rich stars, but this explanation has
yet to be verified. 


The triplet lines are strong and form in the higher photospheric
layers, and thus suffer from non-LTE (NLTE)
effects due to the dilution of each line's source function compared to the Planck function in the line-forming region (Kiselman
2001). The large energy gap between the two lowest energy levels and
levels of higher energy prohibits collisional excitation from maintaining
LTE, and the upper level of the triplet is
underpopulated compared to the lower level (Kiselman 1993). This causes the source function to be
smaller than the Planck function, leading to stronger absorption lines
(Kiselman 1993; Gratton et al.\,1999). Abundances derived from these lines assuming LTE
are thus overestimated. The effect increases as the number of
electrons in the initial (lower) transition state increases, which can
be caused by decreasing gas pressures or increasing temperatures in
the line-forming region, and/or an increase in the number of oxygen
atoms. Thus the discrepancy between LTE and NLTE
calculations and observations is more prominent for hot
($T_{\rm{eff}}\gtrsim 6000$ K) solar-metallicity dwarfs and evolved metal-poor subgiants with decreased surface gravity.  

Multiple groups have prescriptions
for NLTE corrections, which involve establishing the departure from LTE
coefficients ($n_{\rm{NLTE}}$/$n_{\rm{LTE}}$, the ratios of the
populations in NLTE and LTE) from statistical
equilibrium calculations
for varying stellar parameters. 
Takeda (2003) constructs a neutral atomic oxygen
model with 87 levels and 277 radiative transitions, with atomic data
from Kurucz \& Bell\,(1995). In their atomic model, the neutral hydrogen
population is taken
from Kurucz LTE model atmospheres, and the photoionizing radiation is computed
from the same LTE stellar atmospheres, incorporating the line opacity
using Kurucz's (1993) opacity distribution function. The effect of H
I collisions is treated according to Steenbock \& Holweger's (1984)
classicial formula, which is derived from Drawin's (1968) application
of Thomson's theory for electron-atom encounters to collisions between
identical particles. 
Takeda (2003) finds that for a given $T_{\rm{eff}}$, log
$g$, and $\xi$, the NLTE correction to the oxygen abundances is a nearly monotonic function of
EW. They fit the coefficients $a$ and $b$ in their relation
$\Delta=a10^{bW_{\lambda}}$  based on their computed $\Delta$
  values. Here we use this relation and the $a$ and $b$ coefficients
  corresponding to the determined parameters of 55 Cnc to yield
  $\Delta$ corrections to our computed LTE oxygen triplet abundances.   

Ram{\'{\i}}rez et al.\,(2007) compute NLTE corrections using an oxygen model atom with 54
levels and 242 transitions, with atmoic data from
Allende Prieto et al.\,(2003) and fixed temperature and electron
density structures from the Kurucz LTE models. They 
allow the H
and O level populations to depart from LTE by solving rate equations
while recalculating the radiation field with the NLTE stellar
atmosphere code TLUSTY (Hubey \& Lanz 1995), and do not
include H I collisions. Ram{\'{\i}}rez et al.\,(2007) construct a
grid of NLTE abundances directly from curves of growth corresponding
to a range of stellar parameters ($T_{\rm{eff}}$, log $g$, [Fe/H]) and
provide an IDL routine to interpolate within the grid, which we used
here.

Fabbian et al.\,(2009) construct a model atom containing 54 energy
levels and 258 radiation transitions, with atomic parameters from the
NIST Atomic Spectra Database\footnote{http://physics.nist.gov/PhysRefData/ASD/index.html} and radiative and Stark parameters from VALD. They include
fine-splitting of energy levels where appropriate (ground state and
upper level of O I triplet), and the H I collision approximation of
Steenbock \& Holweger\,(1984) scaled by an empirical factor
$S_{H}$, either =0 or =1. Fabbian et al.\,(2009) also include the
most recent electron collision cross sections of Barklem (2007) based
on quantum mechanical calculations; this gives larger NLTE corrections due to
increased intersystem coupling. We obtained their grid of NLTE
corrections and IDL interpolation routine, but it does not cover
[Fe/H]$>$0 or log$N$(O)$>$8.83, so we extrapolated to the
measurements of 55 Cnc. In order to enable direct comparison, we
also interpolated Fabbian et al.'s\,(2009) NLTE corrections to the
same scaling factor as Nissen (2013), $S_{H}=$0.85, which has been
shown to yield the best agreement with observations of O I triplet in
the Sun (Pereira et al.\,2009). 

In Table
\ref{tab:choh} we show the derived
LTE [O/H] abundances from the O I triplet, and also apply the NLTE
corrections of Takeda (2003), Ram{\'{\i}}rez et al.\,(2007), and Fabbian
et al.\,(2009) for comparison. Overall, the NLTE corrections are
between 0.06 and 0.1 dex. 

\subsubsection{Abundance Uncertainties}
There are two components to the uncertainties in derived elemental
abundances -- one from stellar parameter errors and one from the dispersion 
in the abundances derived from different 
absorption lines. To determine the uncertainty due to the stellar parameters, the
sensitivity of the abundance to each parameter was calculated for
changes of $\pm150$ K in $T_{\rm{eff}}$, $\pm0.25$ dex in log $g$, and
$\pm0.30$ km s$^{-1}$ in $\xi$. For the abundances determined through
spectral synthesis, models with this range of stellar parameters were
compared to the data and the elemental abundance adjusted to determine the best fit. The uncertainty due to each
parameter is then the product of this sensitivity and the
corresponding parameter uncertainty. 
The second
uncertainty component 
is 
the uncertainty in the mean, $\sigma_{\mu}$
, for the
abundances derived from the averaging of multiple lines. 
The
total 
uncertainty for each abundance ($\sigma_{\rm{tot}}$) is the quadratic
sum of the three individual parameter uncertainties ($T_{\rm{eff}}$,
log $g$, $\xi$) and
$\sigma_{\mu}$. 

In the case of the O I triplet, the error on [O/H]$_{\rm{NLTE}}$ was
calculated separately for each of the NLTE corrections we applied (see
Table \ref{tab:choh}). For errors on the Ram{\'{\i}}rez et al.\,(2007) and
Fabbian et al.\,(2009) NLTE abundances, we calculated their sensitivity 
to $\pm$150 K $T_{\rm{eff}}$ and $\pm$0.25 dex log $g$. We then
combined these with $\sigma_{\mu}$ for the NLTE
abundances to determine the NLTE abundance errors. Takeda\,(2003) NLTE corrections include a dependence on
$\xi$, so we calculated the sensitivity of these NLTE abundances to
$\xi$ in addition to $T_{\rm{eff}}$ and log $g$, but used changes of $\pm$1 km s$^{-1}$, $\pm$500 K,
$\pm$1.0 dex, respectively, due to the grid spacing of the Takeda\,(2003)
NLTE corrections. As with the other [O/H]$_{\rm{NLTE}}$ errors, we
also included the $\sigma_{\mu}$ for the Takeda\,(2003) NLTE
abundances. 

The final derived stellar parameters and their 1$\sigma$
uncertainties, as well as the derived [Fe/H] 
and [Ni/H] 
values and their 1$\sigma$ uncertanties, are shown in
Table \ref{tab:stellar_params}, along with several
literature values for comparsion. In Table \ref{tab:choh} we detail the
[C/H] and [O/H] values derived from different abundance
indicators. Table \ref{tab:co} shows the range in C/O ratios
 resulting from the different carbon and oxygen abundance
 indicators. These C/O ratios were calculated with the prescription
 log$N_{\rm{55Cnc}}$(O)$=$derived\,[O/H]$_{\rm{55Cnc}}$+log$N_{\odot}$(O)
 and
 log$N_{\rm{55Cnc}}$(C)$=$derived\,[C/H]$_{\rm{55Cnc}}$+log$N_{\odot}$(C), where log$N_{\odot}$(O)$=$8.66 and log$N_{\odot}$(C)$=$8.39 (Asplund
et al.\,2005). The errors on the C/O value are represented by the
quadratic sum of the errors in [C/H] and [O/H].






\section{Results \& Discussion}
The stellar parameters ($T_{\rm eff}$, log g, and [Fe/H]) derived here
compare well with previous determinations in Table
\ref{tab:stellar_params}. The average values from the five literature
sources in Table \ref{tab:stellar_params} are are $T_{\rm eff}$=5268$\pm$38K,
log g=4.45$\pm$0.05, and [Fe/H]=+0.32$\pm$0.03. The differences
between the average values in the literature and those derived here
are, in the sense of `this study - literature', $\Delta T_{\rm eff}$=+82K,
$\Delta$log g=-0.01 dex, and $\Delta$[Fe/H]=+0.02 dex. These differences
are all within the estimated uncertainties presented here and indicate
that there are not large systematic differences between this study and those
published previously.  This result is encouraging, given the challenging
nature of characterizing the relatively rare ``super-metal rich'' stars
with their enhanced line absorption (e.g., Cayrel de Strobel et
al. 1999; 
Taylor 2002; 
Gonzalez \& Vanture 1998; Feltzing \& Gonzalez 2001). Clearly,
55 Cnc is a well-established metal-rich star that happens to be nearby,
hosts a multiple-planet system, and exhibits planetary transits.


Due to its proximity to the Sun and favorable multi-planet geometry, 55 Cnc
is an important object in the study of planet formation, and
thus it is useful to constrain as many of its
fundamental properties as possible. The age of 55 Cnc is uncertain -- 
the $T_{\rm eff}$ or color dependencies
as a function of isochrone age, even for its known metallicity, render
age estimates uncertain by several Gyr's. Ages from $\sim$3-9 Gyr can
fit the position of M$_{\rm V}$ versus $T_{\rm eff}$, or ($V$-$K$), or ($B$-$V$)
isochrones (e.g., Fuhrmann, Pfeiffer \& Bernkopf 1998). Other indicators tend to result in ages from 2-5 Gyr,
such as Eggen's (1985) identification of 55 Cnc as a member of the Hyades
Supercluster with age $\le$2 Gyr. Balinus et al.\,(1997) use the Ca II
K-line activity indicator to estimate an age of 5 Gyr, which is consistent
with their measurement of a rotational activity modulation of 42 days.
Gonzalez (1999) also uses the Ca II K-line to estimate an age of 5 Gyr for
55 Cnc. Taken together, the slow rotation and Ca II K-line suggest a star
perhaps not too different from the Sun in age: almost certainly not younger
than 2 Gyr and probably not much older than 6 Gyr.

Given the 
metal-rich nature of 55 Cnc and the gradual Galactic increase
of C/O with [Fe/H] (Nissen 2013), along with the importance of the natal
C/O ratio in planetary chemistry (e.g., Kuchner \& Seager 2005; Bond et al.
2010), it is important to examine closely the derived C/O ratio in 55 Cnc.
Such scrutiny of C/O takes on added importance when considering the recent
suggestions that some exoplanet host star C/O ratios in the literature 
have been overestimated (Fortney 2012; Nissen 2013).


The forbidden, 
ground-state [O I] 6300.30\,{\AA} line, used in previous host
star studies (e.g., Delgado Mena et al.\,2010; Petigura \& Marcy
2011), gives the lowest oxygen abundance, resulting in the largest
C/O=0.97$\pm$0.31 (using the averaged log$N$(C) of the two [C/H]
indicators). Previous analysis of 55 Cnc using the [O I]
6300.30\,{\AA} line also found a high C/O of 1.12$\pm$0.19 (Delgado Mena
et al.\,2010). Taken at face value, our 6300.30\,{\AA} results would
be 
cosistent with this value within
errors, though allow for 0.66$<$C/O$<$1.27 within 1$\sigma$ uncertainties. 

\begin{figure}[ht!]
\figurenum{1}
\subfigure{ \includegraphics[width=0.5\textwidth]{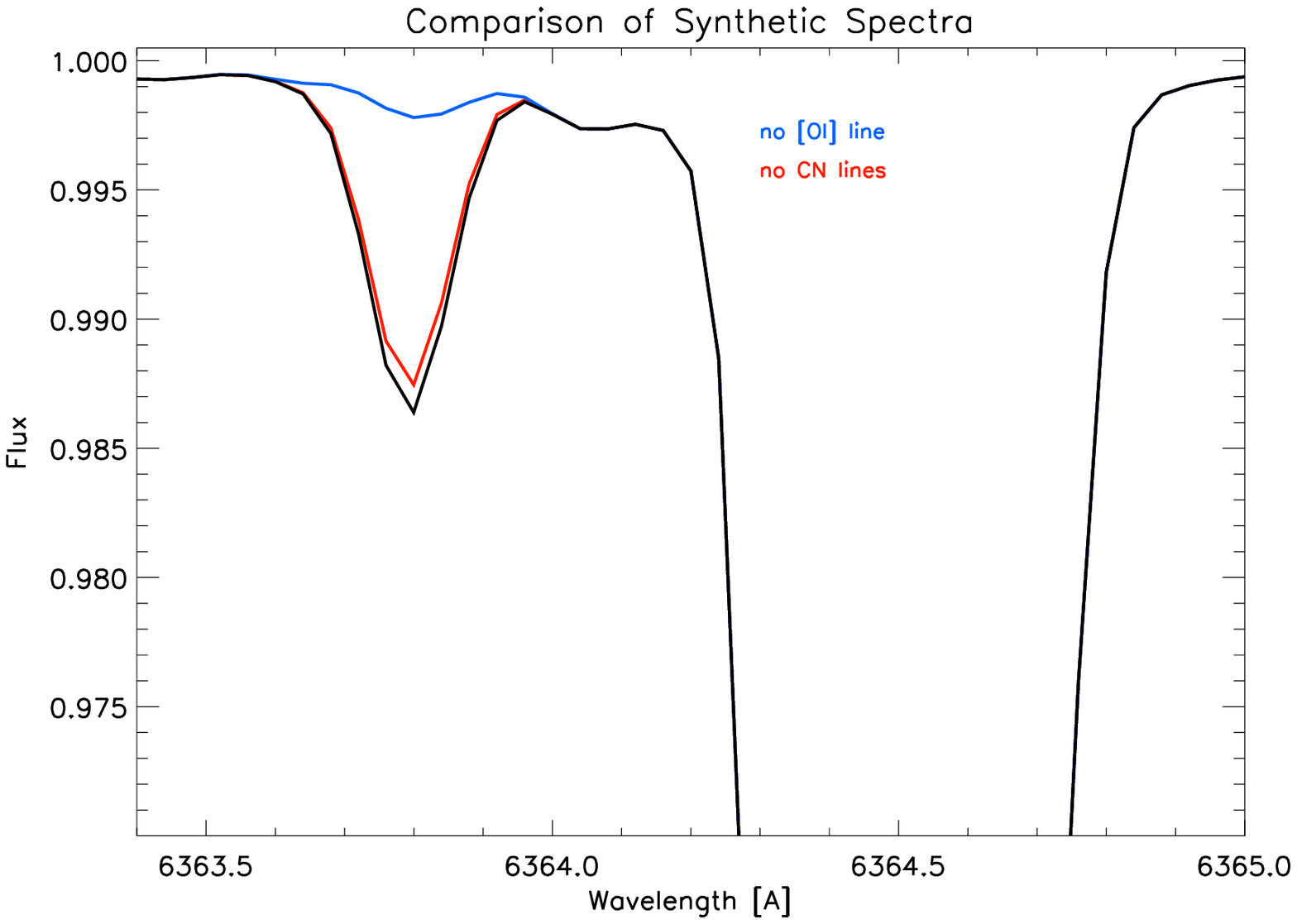}}
\quad
\subfigure{ \includegraphics[width=0.5\textwidth]{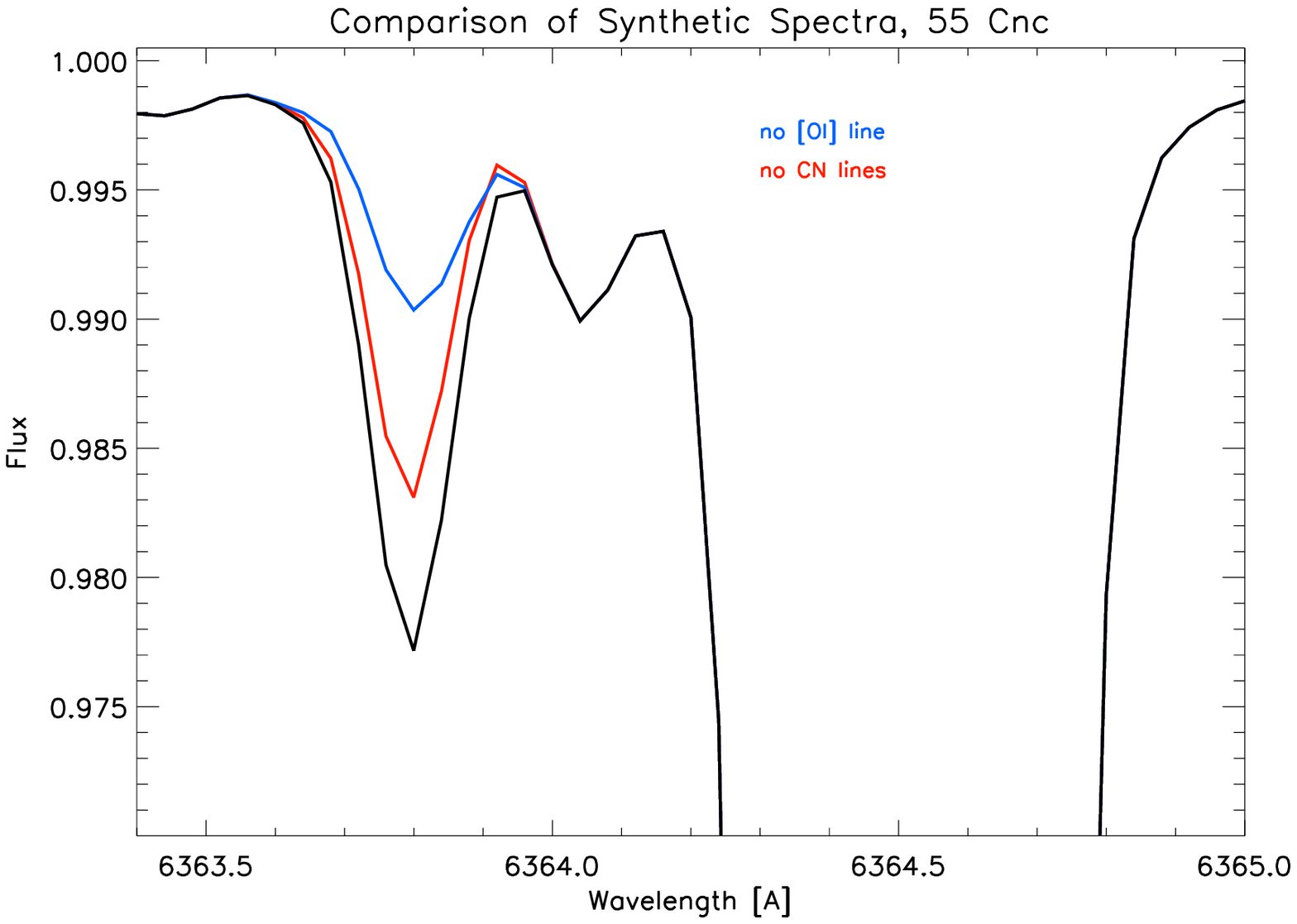}}
\quad
\subfigure{ \includegraphics[width=0.35\textwidth]{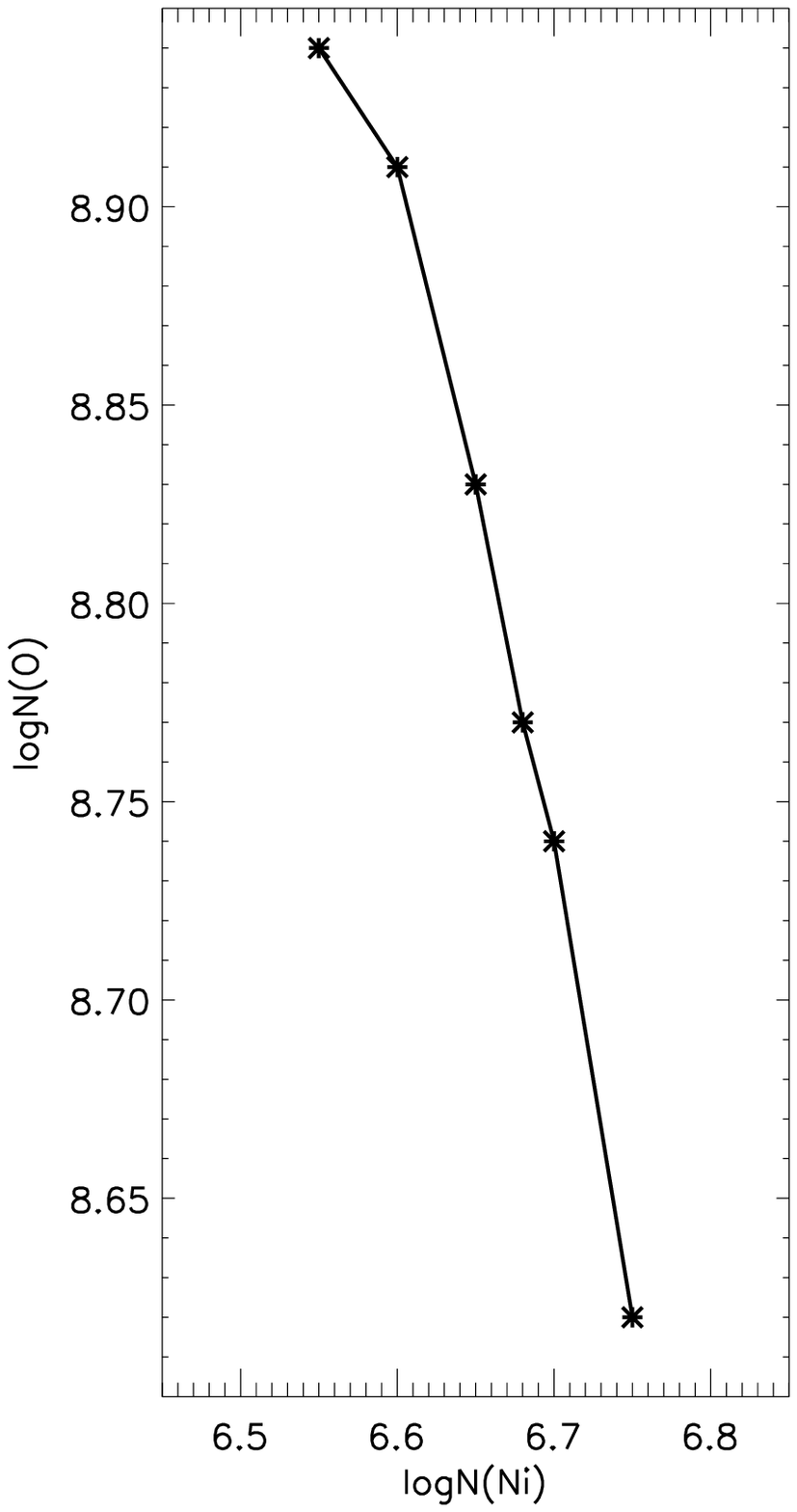} }
 \caption{\textbf{Top row}: The contribution of CN to the blended
   6363\,{\AA} [O I] line is greater in 55 Cnc than in a solar-type star.
The black line shows the model
 spectrum of (left) the sun and (right) 55 Cnc with our adopted stellar paraemters. In red is the stellar model with CN lines removed, showing the contribution of just [O I]. In blue is the
 stellar model with the [O I] line removed, showing the contribution
 of just the CN lines. \textbf{Bottom}: The measured log$N$(O)$_{\rm{55Cnc}}$  changes significantly within the error on the
 determined log$N$(Ni) abundance (log$N$(Ni)$_{\rm{55Cnc}}=$6.68$\pm$0.05).}
\label{fig1}
\end{figure}

However, as noted above, this line is blended with Ni and we find
that in 55 Cnc, the derived [O/H]$_{6300}$ is very sensitive
to the assumed abundance of nickel when performing synthesis analysis. By changing the Ni abundance within our derived error for
[Ni/H] ($\pm$0.05), the best-fit oxygen abundance log$N$(O) varies by
$\sim$0.20 (see Figure \ref{fig1}, bottom). This results in the C/O ratio
varying from $\sim$0.72-1.1, without even considering the 1$\sigma$ \textbf{C/O}
errors (and $\sim$0.42-1.4 considering these errors). 

The [O I] 6363.78\,{\AA} line gives a C/O$=$0.79$\pm$0.23, ranging
within error from $\sim$solar (C/O$_{\odot}$=0.55$\pm$0.10; Asplund et
al.\,2009; Caffau et al.\,2011) to 1. This line is a blend with
CN, which we find contributes a greater amount to the line strength in
the case of 55 Cnc than in the Sun (see Figure \ref{fig1},
top). Additionally, it is weaker than the [O I] 6300\,{\AA} line, and
was found to give higher oxygen abundances in the Sun
(e.g., log$N$(O)$_{6300}$=8.69 vs. log$N$(O)$_{6363}$=8.81), 2 dwarf
stars, and a sub-giant star (Caffau et al.\,2008; Caffau et
al.\,2013). (We find log$N$(O)$_{\odot,6300}$=8.67 vs. log$N$(O)$_{\odot,6363}$=8.84 in our
synthesis analysis of the Sun.) Caffau et al.\,(2013) suggest that the discrepancy is
due to an overestimate in the log $gf$ of the Ni I line that is
blended with the [O I] 6300.30\,{\AA} line. Alternatively, an unknown
blend at 6363\,{\AA} may affect the spectrum of dwarf stars only, as the
6300-6363\,{\AA} discrepancy is not seen in giants (Caffau et
al.\,2013). Certainly this
discussion is still open, and this particular result should be
considered as part of a larger effort to determine [O/H] from both [O I] lines in dwarf
star spectra. Overall, because [O/H]$_{6363}$ for 55 Cnc is
larger, the resulting C/O is smaller than for the 6300\,{\AA} line. 

For the O I triplet at 7771-7775\,{\AA}, the LTE [O/H]$_{\rm{LTE}}=$0.19$\pm$0.17 agrees well with that derived from
the [O I] 6363.78\,{\AA}, 0.17$\pm$0.17, resulting in a similar
C/O$=$0.76$\pm$0.23. As noted, these lines have been shown both theoretically and observationally to
overestimate oxygen abundances in LTE, most significantly at high temperatures and low gravities. We show in Tables \ref{tab:choh} and \ref{tab:co} that three different
NLTE corrections -- Takeda (2003), Ram{\'{\i}}rez et al.\,(2007), and
Fabbian et al.\,(2009) -- give different [O/H] values and 
C/O ratios for 55 Cnc. The corrections are relatively small and, 
perhaps surprisingly, similar despite the different atomic
models, handling of H atom inelastic collisions, and stellar
parameters covered by the corrections. 
For varying NLTE corrections, C/O$_{\rm{55Cnc}}$
ranges from $\sim$0.63-0.70, with a conservative error of $\sim$0.2 based on the LTE
abundances (see \S 2.5). 

However, we note that the validity of applying these NLTE corrections to a cool
and metal-rich star like 55 Cnc is uncertain. With
high-resolution spectroscopy and analysis methods very similar to
those used here, Schuler et al.\,(2004) and (2006) and King \& Schuler
(2005) find a significant $increase$ [O/H]$_{\rm{LTE}}$ values derived from the O I triplet with $decreasing$
$T_{\rm{eff}}$ for dwarfs stars in the Pleiades, M34, 
Hyades open clusters, and the Ursa Major moving group. Such
collections of stars present a unique opportunity for
studying the NLTE effects across stellar temperature and thus mass,
as the stars are presumably, within a single cluster, chemically homogenous and formed at the
same time. This increase in [O/H]$_{\rm{LTE}}$ is in direct contrast with all the NLTE
calculations presented here, which predict negligible effects (e.g.,
$\leq$ 0.05 dex) in dwarfs with $T_{\rm{eff}}\lesssim$5400 K. These
cool cluster dwarf findings are robust, in that the trend remains after re-derivation of
temperatures using different (e.g., photometric) scales, across multiple stellar atmosphere models with or without
convective treatment and varying the mixing-length parameter, and
within all four of these stellar associations. 

The physical mechanism responsible
for the discrepancy in triplet oxygen abundances between calculations
and observations of cool ($T_{\rm{eff}}\lesssim$5400 K)
dwarfs in clusters is not yet certain. By comparing the 
Hyades cluster (600 Myr;
[Fe/H]=+0.13), Pleiades cluster ($\sim$100 Myr,
[Fe/H]=0), and Ursa Major moving group (600 Myr, [Fe/H]=-0.09),
Schuler et al.\,(2006) suggest that the similarity in the observed
[O/H]-$T_{\rm{eff}}$ trend in Hyades and Ursa Major, versus the steeper trend in
Pleiades, points towards an age rather than metallicity effect.
While the
line strengths of the triplet have been shown to increase in a
synthetic solar spectrum when a chromosphere is included (Takeda
1995), Schuler et al.\,(2004) find no correlation between the triplet
[O/H] values and H$\alpha$ and Ca II triplet
chromospheric activity indicators for the Pleiades and M34 stars. This
lack of correlation is confirmed by Schuler et al.\,(2006) between the
Hyades stars' [O/H] and Ca II H+K activity indicators, suggesting that
a more global chromosphere does not contribute to the observed triplet trends in cool cluster
dwarfs. Instead, using simple models including flux contributions to
the triplet region from the quiescent star and both cool and hot spots
is, Schuler et al.\,(2006) are able to reproduce the observed oxygen triplet line strengths in
cool Hyades dwarfs. 
As stellar surface activity is expected to
decrease with age, this result is consistent with the suspected age
dependence of the cool star O I triplet abundances.    

Our derived $T_{\rm{eff}}$ for 55 Cnc (5350$\pm$102 K) places it in
the regime ($T_{\rm{eff}}\lesssim $ 5450 K) where the O I
triplet-temperature trend appears to contradict the canonicial NLTE
oxygen abundance corrections. 
Due to the larger
oxygen triplet NLTE
correction in the Sun versus 55 Cnc, the resulting NLTE-corrected
[O/H] values for 55 Cnc are actually larger than
[O/H]$_{\rm{LTE}}$, although overlap within
errors (see Table \ref{tab:choh}). This behavior is also seen in the cooler stars of Nissen
(2013) -- NLTE corrections in cool stars (even up to $\sim$5660 K)
yield an increase in [O/H]. As a result of higher oxygen abundances,
the C/O ratios for 55 Cnc derived here using the various
[O/H]$_{\rm{NLTE}}$ values are smaller, with a mean of 0.66$\pm$0.07
using the averaged log$N$(C) of the two [C/H] indicators. However, 
all of the stellar associations discussed above are much younger than
the estimated age of 55 Cnc (2-6 Gyr), so the same mechanism(s) may not apply in this case.

Instead of adopting the canonical NLTE corrections, one could estimate
an empirical correction based on the open cluster and moving group
data from Schuler et al.\,(2006). At 5350 K, the T$_{\rm{eff}}$ of 55
Cnc, the typical O I triplet-based abundances are approximately 0.08
dex higher than the mean abundances of the warmer stars in each
cluster. Adopting this difference as a first-order correction, the
resulting O I triplet abundance of 55 Cnc would be [O/H] = 0.11, a value in good agreement with the [O I]-based abundances.



Table \ref{tab:co} presents final abundances and respective C/O ratios from the
individual C I, C$_{2}$, [O I], and O I triplet features. As discussed
earlier, at the temperature and metallicity of 55 Cnc, the 6300\AA\
[O I] feature is dominated by the Ni I blend. Inspection of the O-results
in Table \ref{tab:co} reveals that the 6300\AA\ line yields a lower [O/H]
than the other oxygen indicators. A mean of the O I triplet LTE
and the 6363\AA\ [O I] line
results in log$N$(O)=8.84$\pm$0.01; the 6300\AA\ [O I] abundance falls
significantly outside of this scatter at log$N$(O)=8.74. The decision here, due
to uncertainty caused by significant Ni I blending, is to drop the 
6300\AA\ [O I] result from the final C/O calculation. Additionally,
the various NLTE corrections to the O I triplet abundance may be
unreliable at the temperature and metallicity of 55 Cnc. The
O I triplet NLTE log$N$(O)=8.91$\pm$0.027, different by $\sim$1.9$\sigma$
from the log$N$(O)=8.84$\pm$0.01 calculated from the combined 6363\AA\ [O I] line and O I
triplet LTE values. Therefore we also omit the triplet NLTE results from
the final C/O calculation. We note, though, that including the triplet
NLTE values decreases the mean C/O value only slightly, to
0.71$\pm$0.09, in agreement with the value we choose to report
based on the 6363\AA\ [O I] and O I triplet LTE values. In addition,
including the O I triplet LTE values with the empirical correction
derived from the cool cluster stars increases the mean C/O value
slightly ($\sim$0.03 dex) but is completely consistent with the average we choose to
report here.
 

A final mean C/O ratio is calculated for 55 Cnc based on the 
six values of
C/O in Table \ref{tab:co}, which result from each combination of values from each
respective C and O abundance indicator, excluding those based on the
6300\AA\ [O I] line and the O I triplet NLTE corrections. 
The resulting mean value is
C/O=0.78$\pm$0.08.
Precise values of C/O are important for constraining the composition
of this multiple-planet host star. Several other studies are tackling this
issue with larger samples of mostly giant planet host stars (Delgado Mena
et al.\,2010; Petigura \& Marcy 2011; Nissen 2013). 

\begin{figure}[t]
\figurenum{2}
   \subfigure{\includegraphics[width=0.5\textwidth]{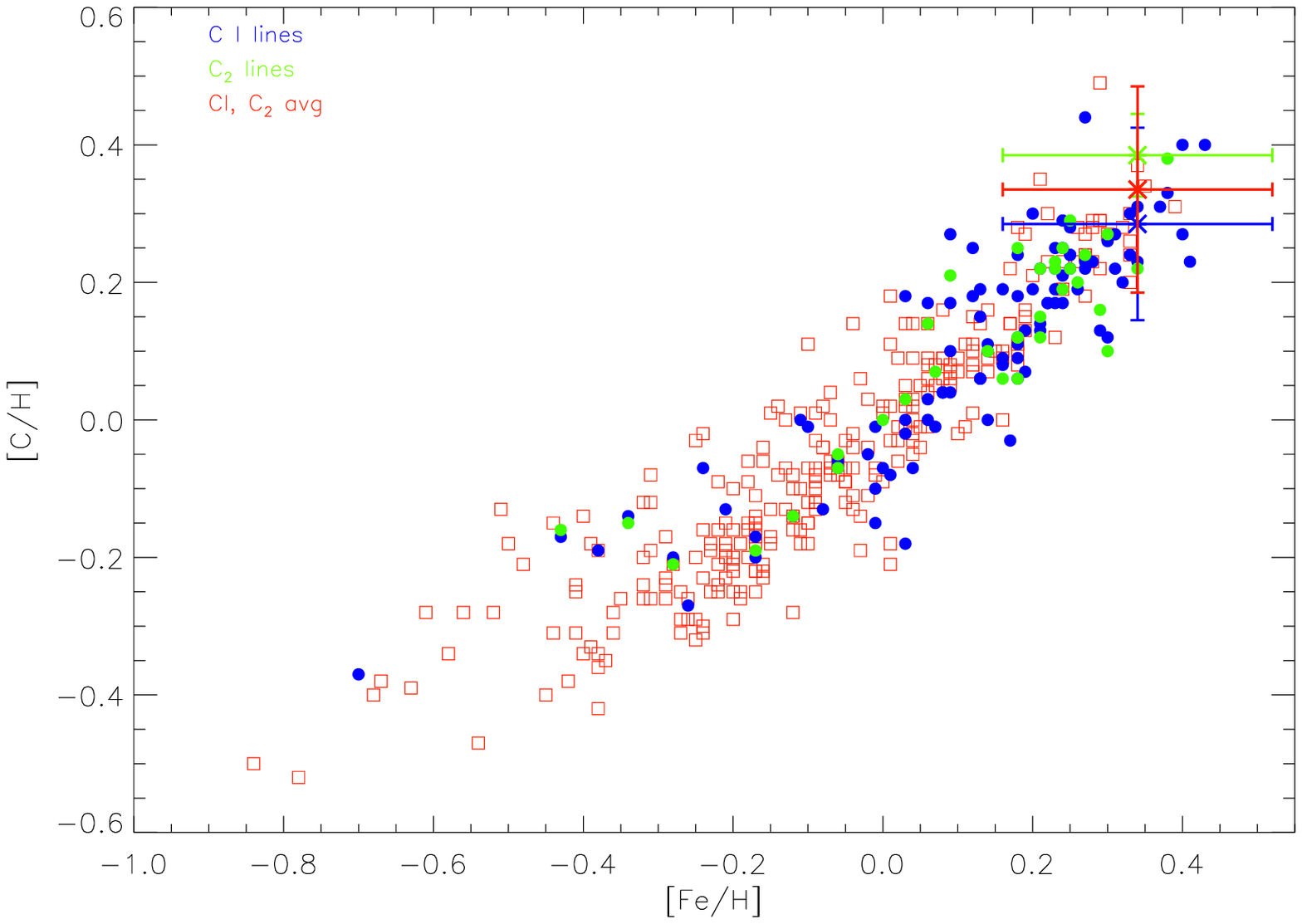}}
\quad
   \subfigure{\includegraphics[width=0.5\textwidth]{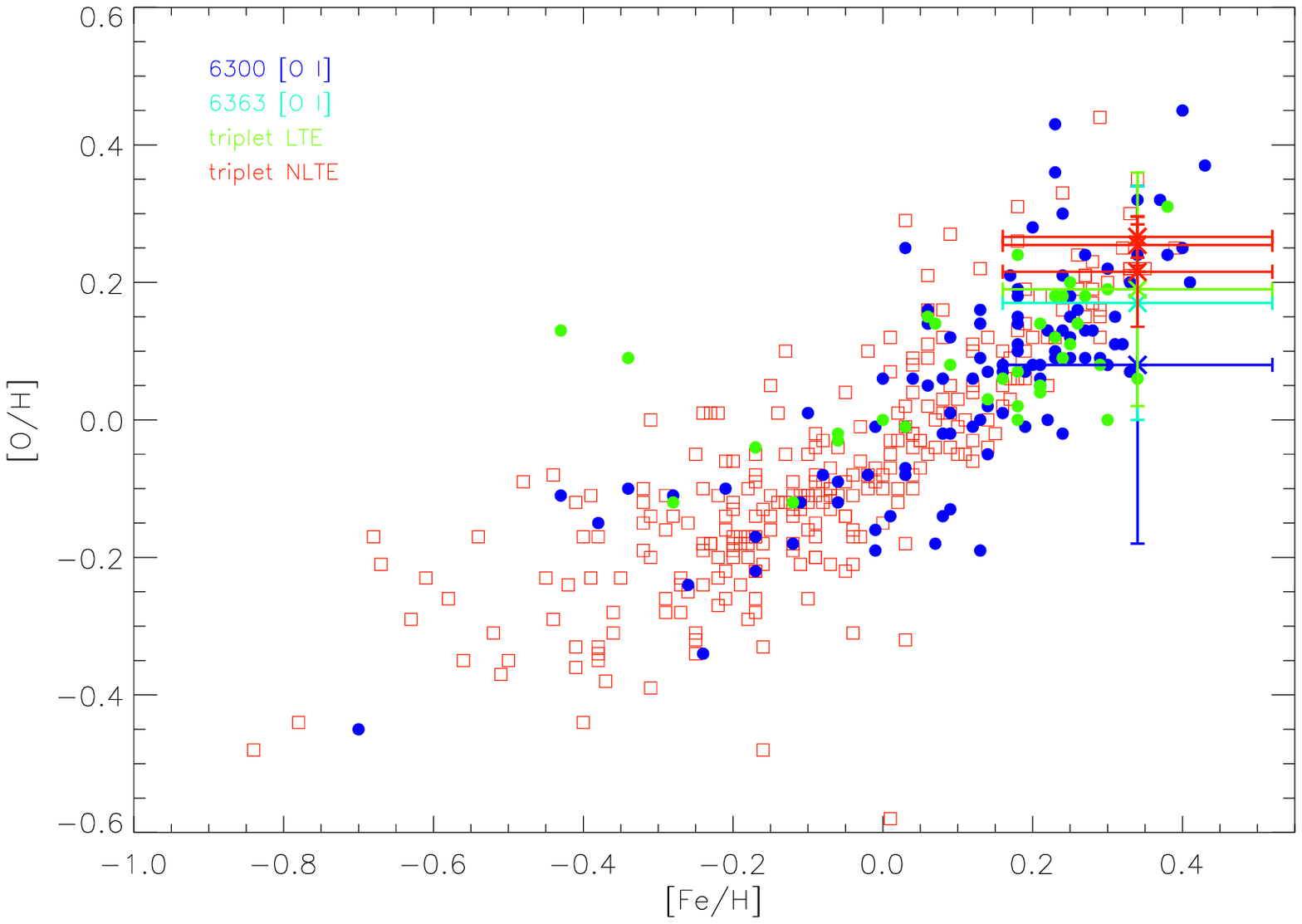}}
\quad
   \subfigure{\includegraphics[width=0.5\textwidth]{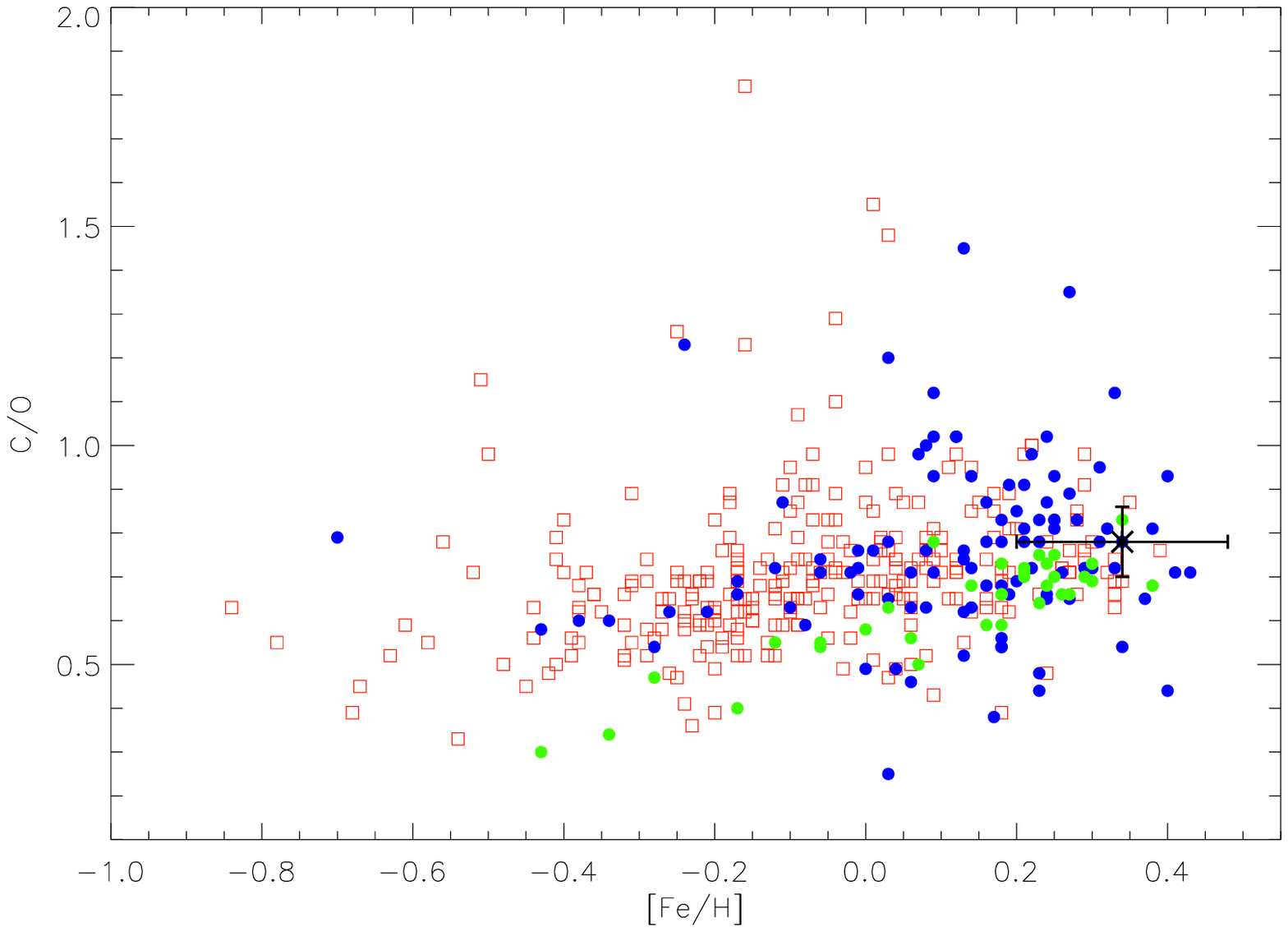}}

\caption{[C/H], [O/H], and C/O versus
  [Fe/H] from Delgado Mena et al.\,(2010) and Nissen\,(2013) [all
  Nissen\,(2013) hosts are in the Delgado Mena et al.\,(2010) host
  sample]. Non-host stars from Delgado Mena et al.\,(2010) are
  plotted with red open squares, while host stars from Delgado Mena
  et al.\,(2010)/Nissen\,(2013) are plotted with blue/green
  circles. Measurements of 55 Cnc from this work are represented by
  large asterisks in each plot (see Table \ref{tab:choh}). In the
  upper plots, we designate measurements from different C and O
  abundance indicators with different colors. Our final mean C/O value
is shown in black in the bottom plot.}
\label{fig2}
\end{figure}

Figure \ref{fig2} shows the values of [C/H], [O/H], and C/O versus [Fe/H] for stars
from the samples noted in the previous paragraph, with the results derived
here for 55 Cnc also shown.  While the spread is still large, the bottom
panel of Figure \ref{fig2} showing C/O versus [Fe/H] indicates that 55 Cnc follows the same 
trends as defined by the larger samples.
With C/O=0.78$\pm$0.08, 55 Cnc
exhibits a ratio that is significantly larger than solar
(C/O$_{\odot}\sim$0.50), but below C/O=1.0 at the 2.75$\sigma$ level. The
value of 0.78 
is lower than the value of C/O=1.12$\pm$0.19
used by Madhusudhan et al.\,(2012) for their carbon-rich models of the
``super-Earth'' exoplanet. 





\section{Conclusions}
The 55 Cnc system 
was the first (Wisdom 2005) and remains one of only a few discovered
systems with five or more planets.
The inner most planet, 55 Cnc e, is one of the most observationally-favorable super-Earth
exoplanets for detailed characterization. 

While previous analyses 
indicate the C/O ratio of 55 Cnc to be
$\geq$1, our analysis indicates that the picture is not so clear. The C/O ratio
of this exoplanet host star is likely 
closer to 
$\sim$0.8. This value is lower than the value adopted by
Madhusudhan et al.\,(2012) in their prediction that the small-mass
exoplanet 55 Cnc e is carbon-rich, and corresponds to the predicted minimum value, $\sim$0.8, necessary to form
abundant carbon-rich condensates, under the assumption of equilibrium
(e.g., Bond et al.\,2010).
Also, possibly the C/O ratio of 55
Cnc's protoplanetary disk was not uniformly identical to its host
star, perhaps causing local carbon enhancements of the gas or grains accreted by 55
Cnc e; carbon-rich planets may still form around oxygen-rich stars
({\"O}berg et al.\,2011; Bond et al.\,2010). Our study places this
system at the theoretically interesting boundary between two diverse
planetary types.  

Measurements of oxygen are challenging in solar-type stars because the oxygen
abundance indicators at optical wavelengths are weak, blended with
other atomic or molecular lines, and/or subject to non-LTE effects. 
Oxygen measurements are even more complicated in cool and high metallicity
stars like 55 Cnc, because of the 
stronger blends
with both atomic and molecular lines, and the uncertainty in NLTE
corrections that do not accurately predict the behavior of line widths
in cool stars.
Our case study demonstrates the 
caution that must be used when
determining exoplanet host star (and any star's) C/O ratios,
particularly the sensitivity of all three major oxygen abundance
indicators to different effects that are not always easy to account
for and change based on stellar parameters. 


\acknowledgements
The authors thank the anonymous referee for her or his helpful
comments. We also wish to recognize and acknowledge the very significant
cultural role and reverence that the summit of Mauna Kea has always
had within the indigenous Hawaiian community. We are most fortunate to
have the opportunity to conduct observations from this mountain. This
research has made use of the Keck Observatory Archive (KOA), which is
operated by the W. M. Keck Observatory and the NASA Exoplanet Science
Institute (NExScI), under contract with the National Aeronautics and
Space Administration. The work of J. T. and C. G. is suppored by NASA's Planetary Atmospheres Program.

{\it Facilities:} \facility{Keck}


\begin{landscape}
\begin{deluxetable}{lcccccc}
\tabletypesize{\scriptsize}
\tablecolumns{7}
\tablewidth{0pc}
\tablecaption{Derived Stellar Parameters and Elemental Abundances for 55 Cnc \label{tab:stellar_params}}
\tablehead{ 
\colhead{Parameter} & \colhead{this work$^{a}$}  & \colhead{Valenti \&
  Fischer 2005$^{b}$} & \colhead{Butler et al.\,2006} & \colhead{Ecuvillon
  et al.\,2004 or 2006}  &  \colhead{Takeda et al.\,2007} & \colhead{Zielinski et al.\,2012}}
\startdata
$T_{\rm{eff}}$ (K) & 5350$\pm$102  & 5235$\pm$15 &
5235$\pm$44  & 5279$\pm$62 & 5327$\pm$49 & 5265$\pm$15  \\

log $g$ (cgs) & 4.44$\pm$0.30     & 4.45$\pm$0.02
& 4.45$\pm$0.06 & 4.37$\pm$0.18 & 4.48$^{-0.01}_{+0.05}$  & 4.49$\pm$0.05 \\

$\xi$ (km s$^{-1}$)& 1.17$\pm$0.14 & \nodata &
\nodata    & 0.98$\pm$0.07 & \nodata &\nodata \\

[Fe/H] & 0.34$\pm$0.18  & 0.31$\pm$0.01 &0.32$\pm$0.03 & 0.33$\pm$0.07
& 0.37$\pm$0.04 & 0.29$\pm$0.07 \\


[Ni/H] & 0.43$\pm$0.05  & 0.37$\pm$0.01 & \nodata & 0.39$^{c}$ &
\nodata & \nodata   \\
\enddata

\tablenotetext{a}{Adopted solar parameters: $T_{\rm{eff}}=$5777 K, log $g=$4.44, and $\xi=$1.38 km s$^{-1}$.}
\tablenotetext{b}{Uncertainties from fitting a single ``standard'' star, divided by
  $\sqrt{n}$, where $n=8$, the number of observations of 55 Cnc in
  Valenti \& Fischer (2005).} 
\tablenotetext{c}{Derived by Delgado Mena et al.\,(2010) with spectra from the
  CORALIE survey, using Ecuvillon stellar parameters. Specific errors not provided.} 
\end{deluxetable}
\end{landscape}

\begin{deluxetable}{lccccccc}
\tablecolumns{8}
\tablewidth{0pc}
\tabletypesize{\scriptsize}
\tablecaption{Lines Measured, Equivalent Widths, and Abundances \label{tab:lines}}
\tablehead{ \colhead{Ion} & \colhead{$\lambda$} & \colhead{$\chi$} & \colhead{log $gf$} & \colhead{EW$_{\odot}$} & \colhead{log$N_{\odot}$} & \colhead{EW$_{\rm{55Cnc}}$} &  \colhead{log$N_{\rm{55Cnc}}$}\\ 
  \colhead{ } & \colhead{({\AA})} & \colhead{(eV)} & \colhead{} & \colhead{(m{\AA})} & \colhead{ } & \colhead{(m{\AA})} & \colhead{} }
\startdata
C I & 5052.17 & 7.68 & -1.304 & {33.7}  &{8.45}  & {31.3}&{8.72}  \\

{ } & 5380.34 & 7.68 & -1.615 & {20.7} & {8.48} & {20.6} &{8.78} \\  

$[$O I$]$ & 6300.30 & 0.00 & -9.717 & {5.6}&{8.67$^{a}$} &{7.2} & {8.75$^{a}$} \\

{} & 6363.79 & 0.00 & -10.185 &{1.6} &{8.84$^{a}$} &{3.4} & {9.01$^{a}$} \\

O I & 7771.94 & 9.15 & 0.369 &{69.6} &{8.83} &{48.9} & {9.00} \\

{} & 7775.39 & 9.15 & 0.001 &{46.8} &{8.81} & {33.6}& {9.01} \\
\enddata
\tablenotetext{a}{Abundance derived through synthesis analysis.}
\tablenotetext{b}{\,LTE abundance.}
\tablecomments{This table is available in its entirety in a machine-readable form in the online journal. A portion is shown here for guidance regarding its form and content.}
\end{deluxetable}

\begin{landscape}
\begin{deluxetable}{lcccc}
\tabletypesize{\scriptsize}
\tablecolumns{5}
\tablewidth{0pc}
\tablecaption{55 Cnc Carbon and Oxygen Abundances from Different Indicators \label{tab:choh}}
\tablehead{ 
\colhead{Abundance Indicator} & \colhead{this work$^{a}$}
&\colhead{Ecuvillon et al.\,2004 or 2006$^{b}$} & \colhead{Delgado Mena et al.\,2010$^{c}$} & \colhead{Petigura \& Marcy 2011$^{d}$} }
\startdata
[C I/H] &  0.29$\pm$0.14 &0.31$\pm$0.10    & 0.30 & 0.13$\pm$0.06\\

[C$_2$/H] & 0.39$\pm$0.06 & \nodata   & \nodata & \nodata \\

[O/H]$_{6300}$ & 0.08$\pm$0.26 &  0.13$\pm$0.11 & 0.07 & \nodata \\

[O/H]$_{6363}$ & 0.17$\pm$0.17 &  \nodata  & \nodata & \nodata \\

[O/H]$_{7772, 7775; \rm{LTE}}$ & 0.19$\pm$0.17 & 0.21$_{\rm{no\,LTE\,error\,given}}^{\rm{avg\,of\,all\,triplet\,lines}}$  & \nodata & \nodata \\

[O/H]$_{7772, 7775; \rm{NLTE}}$ & 0.22$\pm$0.08$_{\rm{Takeda\,NLTE}}$&0.03$\pm$0.11$_{\rm{Ecuvillon\,NLTE\,model}}^{\rm{avg\,of\,all\,triplet\,lines}}$  & \nodata & \nodata \\
                        & 0.25$\pm$0.03$_{\rm{Ramirez\,NLTE}}$  &    &         & \\
                       &0.27$\pm$0.03$_{\rm{Fabbian\,NLTE}}$  &    &         & \\
\enddata
\tablecomments{The NLTE corrections calculated from Fabbian et
  al.\,(2009) have been interpolated to a Drawin formula scaling factor S$_{\rm{H}}=0.85$, as in
  Nissen 2013.}
\tablenotetext{a}{ The $\pm$ errors here $=$ the final combined abundance
  uncertainties due to both stellar parameters and (if applicable) the dispersion in
  abundances derived from multiple lines. Errors on [O/H]$_{7772,
    7775; \rm{NLTE}}=$ the uncertainties due to the stellar parameters
  that factor into in each
  of the NLTE calcuations ($T_{\rm{eff}}$ and log $g$ for
  Ram{\'{\i}}rez and Fabbian; $T_{\rm{eff}}$, log $g$, and $\xi$ for
  Takeda) and the dispersion in the derived NLTE abundances.} 
\tablenotetext{b}{The $\pm$ errors here factor in uncertainties in
  stellar parameters, continuum determination, and (if applicable)
  the standard deviation of multiple measured lines.}
\tablenotetext{c}{Specific errors not provided.}
\tablenotetext{d}{The $\pm$ error $=$15\% and 85\% confidence limits.}
\end{deluxetable}
\end{landscape}

\begin{landscape}
\begin{deluxetable}{lccccccc}
\tablecolumns{8}
\tabletypesize{\scriptsize}
\tablewidth{0pt}
\tablecaption{C/O Ratios of 55 Cnc Based on Different C and O
  Abundance Indicators\label{tab:co}}
\tablehead{
\colhead{} & \colhead{log$N$(O)} & \colhead{6300\,{\AA} [O I]} &
\colhead{O I triplet} & \colhead{O I triplet} &
\colhead{O I triplet} & \colhead{O I triplet} & \colhead{6363\,{\AA} [O I]}\\ 

\colhead {} & \colhead{} & \colhead{log$N$(Ni)=6.68} & \colhead{LTE} &
\colhead{NLTE Takeda\,(2003)}
& \colhead{NLTE Ram{\'{\i}}rez et al.\,(2007)} &
\colhead{NLTE Fabbian et al.\,(2009)} & \colhead{}
}
\startdata
log$N$(C)                        & {}    & 8.740 & 8.845&8.875 & 8.915 & 8.926 & 8.830 \\
\hline
two blue C I lines& 8.675 & 0.861$\pm$0.299 &
0.676$\pm$0.217& 0.631$\pm$0.217 & 0.576$\pm$0.217 & 0.561$\pm$0.217 &
0.700$\pm$0.217 \\
\hline
two C$_2$ lines & 8.775 & 1.084$\pm$0.272 &
0.851$\pm$0.178& 0.794$\pm$0.178 & 0.725$\pm$0.178 & 0.706$\pm$0.178 &
0.881$\pm$0.178 \\
\hline
C I and C$_2$ averaged & 8.725 & 0.966$\pm$0.306 &
0.759$\pm$0.226 & 0.707$\pm$0.226 & 0.646$\pm$0.226 & 0.629$\pm$0.226
& 0.785$\pm$0.226\\
\enddata
\tablecomments{The log$N$(O or C) values are calculated as
  [X/H]$+$log$N_{\odot}$(X), with log$N_{\odot}$(O)=8.66 and
  log$N_{\odot}$(C)=8.39 (Asplund et
  al.\,2005). The $\pm$ errors here $=$ the final combined abundance
  uncertainties due to both stellar parameters and (if applicable) the dispersion in
  abundances derived from multiple lines. We adopted the more
  conservative (larger) [O/H]$_{\rm{LTE}}$ errors for the
  [O/H]$_{\rm{NLTE}}$ values. Serendipitously, the
  [O/H]$_{\rm{LTE}}$ error $=$ the [O I]$_{6363}$ error; hence,
  columns 4-8 have identical errors.}
\end{deluxetable}
\end{landscape}
\clearpage

\end{document}